\documentclass[aps,pra,showpacs,twocolumn,superscriptaddress,notitlepage,nofootinbib]{revtex4-1}
\usepackage{mathtools}
\usepackage{amsfonts}
\usepackage{latexsym}
\usepackage{relsize}
\usepackage{euscript}
\usepackage{amssymb}
\usepackage{graphicx}
\usepackage{amsmath}
\usepackage{amsbsy}
\usepackage{amsthm}
\usepackage{bbm}
\usepackage{bm}
\usepackage{epsfig}
\usepackage{epstopdf}
\usepackage{dsfont}

\usepackage[colorlinks]{hyperref}
\usepackage[figure,table]{hypcap}
\usepackage{enumerate}
\usepackage{geometry}
\hypersetup{
	bookmarksnumbered,
	pdfstartview={FitH},
	citecolor={darkgreen},
	linkcolor={darkred},
       linktoc={page},
	urlcolor={darkblue},
	pdfpagemode={UseOutlines}}

\usepackage{color}

\definecolor{darkgreen}{RGB}{40,130,40}
\definecolor{darkblue}{RGB}{0,0,190}
\definecolor{darkred}{RGB}{238,0,0}
\usepackage{soul}
\usepackage{float}
\usepackage{algorithm}

\floatname{algorithm}{Protocol}

\newcommand{\djj}{d\kern-0.4em\char"16\kern-0.1em}

\newtheorem{prop}{Proposition}\def\PRO{\begin{prop}}\def\ORP{\end{prop}}
\newtheorem{coro}{Corollary}\def\COR{\begin{coro}}\def\ROC{\end{coro}}
\newtheorem{theo}{Theorem}\def\TH{\begin{theo}}\def\HT{\end{theo}}
\def\TH{\begin{theo}}\def\HT{\end{theo}}
\newtheorem{defi}[prop]{Definition}\def\DE{\begin{defi}}\def\ED{\end{defi}}
\newtheorem{lemme}[prop]{Lemma}\def\LE{\begin{lemme}}\def\EL{\end{lemme}}

\newtheorem{wlemme}[prop]{I wish I had this Lemma}\def\wL{\begin{wlemme}}\def\Lw{\end{wlemme}}

\newtheorem{wwlemme}{I wish I had this Theorem}\def\wT{\begin{wwlemme}}\def\Tw{\end{wwlemme}}

\begin{document}

\title{Quantum Google Algorithm: Construction and Application to Complex Networks}
\author{G.D. Paparo}
\affiliation{Departamento de F\'isica Te\'orica I, Universidad Complutense, 28040 Madrid, Spain.}
\author{M. M\"uller}
\affiliation{Departamento de F\'isica Te\'orica I, Universidad Complutense, 28040 Madrid, Spain.}
\author{F. Comellas }
\affiliation{Departament de Matem\`atica Aplicada IV, Universitat Polit\`ecnica de Catalunya, 08034 Barcelona, Spain.}
\author{M.A. Martin-Delgado}
\affiliation{Departamento de F\'isica Te\'orica I, Universidad Complutense, 28040 Madrid, Spain.}

 \begin{abstract}
We review the main findings on the ranking capabilities of 
the recently proposed Quantum PageRank algorithm~\cite{gdpmamd2011,paparo2013quantum} applied to large complex networks. 
The algorithm has been shown 
to identify unambiguously  the underlying topology of the network and  to be capable of clearly highlighting the structure of secondary hubs of networks. Furthermore, it can resolve the degeneracy in importance of the low lying part of the list of rankings. 
Examples of applications include real world instances from the WWW, which typically display a scale-free network structure and models of hierarchical networks.
The quantum algorithm has been shown to display an increased stability with respect to a variation of the damping parameter, present in the Google algorithm, and a more clearly pronounced power-law behaviour in the distribution of importance among the nodes, as compared to the classical algorithm. 
 \end{abstract}

%
\maketitle
\section{Introduction}
\label{sect:Intro}

One of the most important 
revolutions in the world of communication that have emerged in the last decades is the World Wide Web (WWW).
Its presence in modern  life is ubiquitous and its use in the daily life has made many tasks easier. 
However, the potential power offered by the web can only be leveraged if there are effective methods to search, retrieve and order information. 
This ability is inextricably connected to the problem of organizing and ranking the information in the WWW.

Various approaches have been set forward to bring order in the information stored. The most effective strategy has been to think of the information stored in elementary units (nodes) which are connected and form a network. Therefore, with this mindset, the importance or relevance of a node depends on the way it is connected to the other nodes in the network and the hyperlink structure becomes a key ingredient. 
S. Brin and L. Page based PageRank~\cite{brin1998anatomy,page1999pagerank,marijuan11,shepelyansky} on this idea. The algorithm lies at the core of Google's well-known and widespread search engine, by means of which to date by far the largest portion of web searches are carried out.

The assessment of the importance of a node allows the search engine to retrieve the information in the 
node, which corresponds to a webpage in a static WWW, and to output
what is stored in the most relevant nodes. It is therefore crucial to be able to rank nodes in a network.
Google's search algorithm provided a fast, efficient and objective way of ranking information as the content was being evaluated with respect to its relation to the rest. On the opposite, other approaches used by competing search engines, were to retrieve information by database search to reply to user queries. In this case, the choice of which information would better satisfy the user was made according to subjective criteria.

Recently, a new computational framework, quantum computation~\cite{NC,rmp}, has made its appearence. In the past decades quantum mechanics has been demonstrated to be advantageous when performing computational tasks and the upshots include speedups, i.e. that fewer computational steps are needed to reply to the same question or solve the same problem. 

Database search is a prominent example from the class of problems, where quantum computation has shown to display a speedup over the best known classical search algorithms. This speedup has been attained with Grover's algorithm~\cite{grover1996fast} and variants thereof~\cite{galindo2000family}. 
More recently, also artificial intelligence~\cite{paparo2014quantum} has benefited from the application of ideas from quantum computation.

Small versions of quantum computers have been built and many of the components have been demonstrated in proof-of-principle experiments~\cite{ladd2010quantum}
and attempts to scale these  systems up are under way.
It is therefore very tempting to take advantage of quantum mechanics for the problem of ranking. In~\cite{gdpmamd2011} we proposed a quantum algorithm  to rank nodes in a network. We showed how quantum dynamics can help in accomplishing the task of ranking, thus providing a more detailed and complete evaluation of the importance of nodes in a network. 
In~\cite{paparo2013quantum} we extended the study of the features and performance of the algorithm when applied to large complex networks.

In classical information processing the concept of networking, i.e. connecting various computing devices, has been shown to be advantageous, the WWW being born from this idea. Similarly, also in quantum computing the possibility of establishing networks has been considered. Indeed, quantum networks, where information can be stored in quantum degrees of freedom, have been proven to allow theoretically provably secure quantum cryptography~\cite{bennett1984quantum,ekert1991quantum}. At the present stage, quantum networks are under active investigation and some early versions of them have been
designed and implemented in recent years~\cite{darpa,SECOQC,UQCC,SwissQuantum,vicente1,ETSI}.
There are other more demanding proposals for quantum networks~\cite{kimble2008quantum,wiersma2010random,luo2014efficient} 
based on entanglement connections
that need quantum repeaters~\cite{briegel1998quantum,dur1999quantum,sangouard2011quantum} in order to 
be able to reliably transmit quantum information over large distances in the presence of losses or noise~\cite{lauritzen2011approaches,simon2010quantum,lauritzen2010telecommunication}. 
Another alternative to build different types of quantum networks makes use of quantum percolation protocols~\cite{acin2007entanglement,perseguers2008entanglement,wei2013renormalization,cuquet2009entanglement,cuquet2011limited}.
In the future, it will be natural to store information in these networks and the problem of bringing order to this information will become of paramount importance. It is therefore tantalizing to think about a way of ranking and organising this information.
With our Quantum PageRank algorithm we have shown a route to accomplish such  a task.
Building a quantum network has been targeted as a fundamental goal in quantum information. 
Most likely, it will be a more feasible goal in the near future than the construction of the first fully fledged quantum computer. 
Therefore, it is highly likely that quantum networks where information is stored in quantum degrees of freedom will be realised before large scale quantum computers become available.
It is therefore desirable to have an algorithm that can be efficiently implemented also on a classical computer. The availability of such an algorithm will allow one to work in such a scenario where a quantum network is present but a large scale quantum computer is not.

Whether they are quantum or classical, networks are a central concept in the analysis of information. Indeed, the idea of determining the importance of atoms of information (nodes) with respect to the others renders the connectivity structure a central ingredient for the problem of ranking.
It is not hard to understand why this way of looking at the problem has greatly fostered the field of complex network science. The scientific community has devoted a considerable effort to analyse and understand networks like the WWW, trying to uncover the reasons behind its connectivity structure, with the hope of further leveraging its power. The upshots of this field of research have not been limited to this area but have shed light on the functionality of different types of complex  networks, ranging from transportation and communication to biological and social networks.
Starting with the seminal papers by Watts and
Strogatz on small-world networks~\cite{small-world} and by Barab\'asi and
Albert on scale-free networks~\cite{Barabasi_Albert_99}, researchers realised that many real-world networks belong to a class  known as small-world scale-free networks. These networks exhibit both strong local clustering, {\it i.e.} nodes have many mutual neighbours, and a small average path length while sharing another important characteristic: the number of links of nodes usually obeys a power-law distribution, {\it i.e.} the network is scale-free. Moreover, it has  been found that many real networks, including the WWW,  are also self-similar, see \cite{Song_2005}.
Such properties can often be related to a modular and hierarchical structure and organisation which is 
essential for their communication and dynamical processes~\cite{RaBa03,Barabasi-Oltvai-04,BaDeRaYoOl03}. 
On the other hand, this hierarchical structure could explain the existence of nodes 
with a relatively large number of links or hubs, which play a critical role in the information flow of the system. Hubs are also associated with a low average distance in the network. Several general reviews and books on complex networks  are now available, to which we refer the reader interested in more information on this topic~\cite{statistical2,Newman-Nets,CoHa10}.

{\it Outline and main aspects.} In this paper we review the Quantum PageRank algorithm~\cite{gdpmamd2011,paparo2013quantum} and illustrate the main findings when applied to complex networks of different sizes and topologies, like the scale-free and hierarchical ones. 
In sect.~\ref{sect:one} we recapitulate Google PageRank's algorithm and introduce Quantum PageRank. We show the main features found already in the analysis of small networks. In sect.~\ref{sect:3} we show that the algorithm: i/ is able to better rank nodes, highlighting the structure of hubs and secondary hubs on scale-free networks, ii/ that it is able to recognize the hierarchical structure in networks and to amplify the importance difference due to the local (intra-layer) connectivity structure.
A necessary condition for a ranking algorithm to rank in scale-free networks is that the diffusion phenomenon on networks tends to localise the walker. In sect.~\ref{sect:Further_Aspects} we review our results on the iii/ localisation properties on the networks.
The output, {\it i.e.} the rankings, greatly depends on the damping parameter, put by hand in the classical PageRank algorithm to ensure convergence. 
It is therefore interesting to assess to which extent the quantum algorithm's output depends on this parameter. In sect.~\ref{sect:Further_Aspects} we review that  iv/  it is more robust with respect to the variation of this parameter and that its dependence is milder in the quantum algorithm, thus solving the problem of the arbitrariness with which this parameter was tuned.
In this section we will also illustrate v/ the origin of the increased visibility of the network structure analysing  the scaling behavior of the rankings. As we will see, the ability of the quantum algorithm to highlight more hubs relies on a more homogenous distribution of importance among nodes. Indeed, classical PageRank tends to give a very high importance to the main hubs thus putting on an equal footing all the other nodes. 
Finally, we will also briefly discuss vi/ the sensitivity of the quantum algorithm with respect to coordinated attacks of the most important nodes in scale-free networks.

\section{From Classical to Quantum PageRank}
\label{sect:one}


During the last decade of the past century the web was growing at a pace that began to outstrip the ability of existing search engines to yield useable results. Furthermore, existing approaches relied on database search which was based on subjective criteria. In the second part of the 1990's~\cite{wiki2014Timeline} many researchers proposed solutions in order to make web searches effective and as objective as possible. During the search for alternative approaches the role of the hyperlink structure in the web was recognized and this led to the introduction of new algorithms, as e.g. Hypersearch \cite{marchiori1997quest}, whose main idea was to assign the importance of a webpage in relation to the ones linking to it. 
Brin and Page founded Google in 1998, endowing its search  with PageRank's algorithm. The construction of the latter ranking protocol was influenced by the debate on the role of the hyperlink structure on ranking. Therefore a major distinction between PageRank (PR)  and previous approaches used by other search engines is the fact that PR had an objective character, while others were based on subjective criteria of the contents of the pages, because they were built as a collection of links that people in companies  stored on a regular basis. 
In other words, PR is dynamical while the other approaches are static and subjective with respect to the contents of the pages. 

The way most search engines, including Google, work is to continually retrieve pages from the web, index the words in each document, and store this information. Each time a user asks for a web search using a search phrase, such as ``word", the search engine outputs  all the pages on the web that contain ``word'' or are semantically related to it. 
 
A problem arises naturally: Google now claims to index 50  billion pages~\cite{web2014websize}. Roughly 95 \% of the text in web pages is composed from a mere 10,000 words. This means that, for most searches, there will be a huge number of pages containing the words in the search phrase. 
It is clearly of paramount interest to be able to rank these pages according to their relevance so that the pages can be sorted with the most important pages at the top of the list. Google's success is largely due to PageRank's effectiveness in ranking pages in the WWW.

\subsection*{Google PageRank}

The key idea of Google's PageRank algorithm is that the importance of a page is given by how many pages link to it. Moreover, the contribution of each page is larger if the number of pages it links to is smaller.
Defining $I(P_i)$ as the importance of a page $P_i$ and $B_i$ as the set of pages linking to it, the aforementioned idea can be put into an equation as follows:
\begin{equation}
I(P_i):= \sum_{j \in B_i} \frac{I(P_j)}{\rm{outdeg}(P_j)},
\label{ eq: definition of pagerank1}
\end{equation}
where outdeg($P_j$) is the outdegree (i.e. the number of outgoing links) of the page $P_j$.
Let us take into account the hyperlink -or connectivity- matrix, defined as follows:
\begin{equation}
H_{ij} := 
\begin{cases} 
\;1/\rm{outdeg}(P_j)   \, & \mbox{if }P_j \in B_i  \\
\;0  \, &  \mbox{otherwise }
\end{cases}
\label{eq:definition hyperlik matrix}
\end{equation}
and consider the vector $I$ whose components are the PageRanks $I(P_i)$. We can state the problem of finding the importance of a page as the problem of solving the self-consistency equation:
\begin{equation}
I = H I 
\label{ eq: definition of pagerank1 matrix form}
\end{equation}
It is apparent that the problem of finding PageRanks is the equivalent of finding the eigenvector with eigenvalue one of a matrix.
The hyperlink matrix $H$ is a square matrix whose dimension is given by the number of webpages indexed by Google. This means that $H$ is  a  $50$  billion by $50$  billion matrix and therefore, it is a very tough challenge to solve the eigenvalue problem. However, $H$ is  sparse in general, i.e. most of the entries in $H$ are zero; in fact, studies~\cite{langville2004deeper} 
 show that webpages, on average, have ten hyperlinks, meaning that all but ten entries, on average, in every column are zero. The sparseness of the hyperlink matrix is a key feature to ensure that a seemingly formidable problem in theory is tractable in practice.
 
The power method is one of the workhorses used to solve such a problem and find the stationary vector $I$ of the matrix $H$. It is an iterative method that starting from a candidate vector $I_0$ produces a sequence of vectors $I_k$ by applying iteratively the matrix $H$ {\it i.e. }:
\begin{equation}
I^{k+1} = H I^{k} 
\label{ eq: definition of pagerank1 power method2}
\end{equation}
However, formulating in such a way the problem of finding the PageRank vector will not output, in general, a meaningful vector. Indeed, the sole use of the connectivity structure in the procedure, although necessary, is not sufficient. The power method might not converge  at all or, if it does, the stationary vector's components cannot be interpreted as the webpages' importances. In the following we will go through the modifications to this idea to ensure that we obtain a sensible answer to the problem of ranking.

\subsection*{Patching the Algorithm}

\begin{figure}
\centering
\includegraphics[keepaspectratio=true,width=.95\linewidth]
{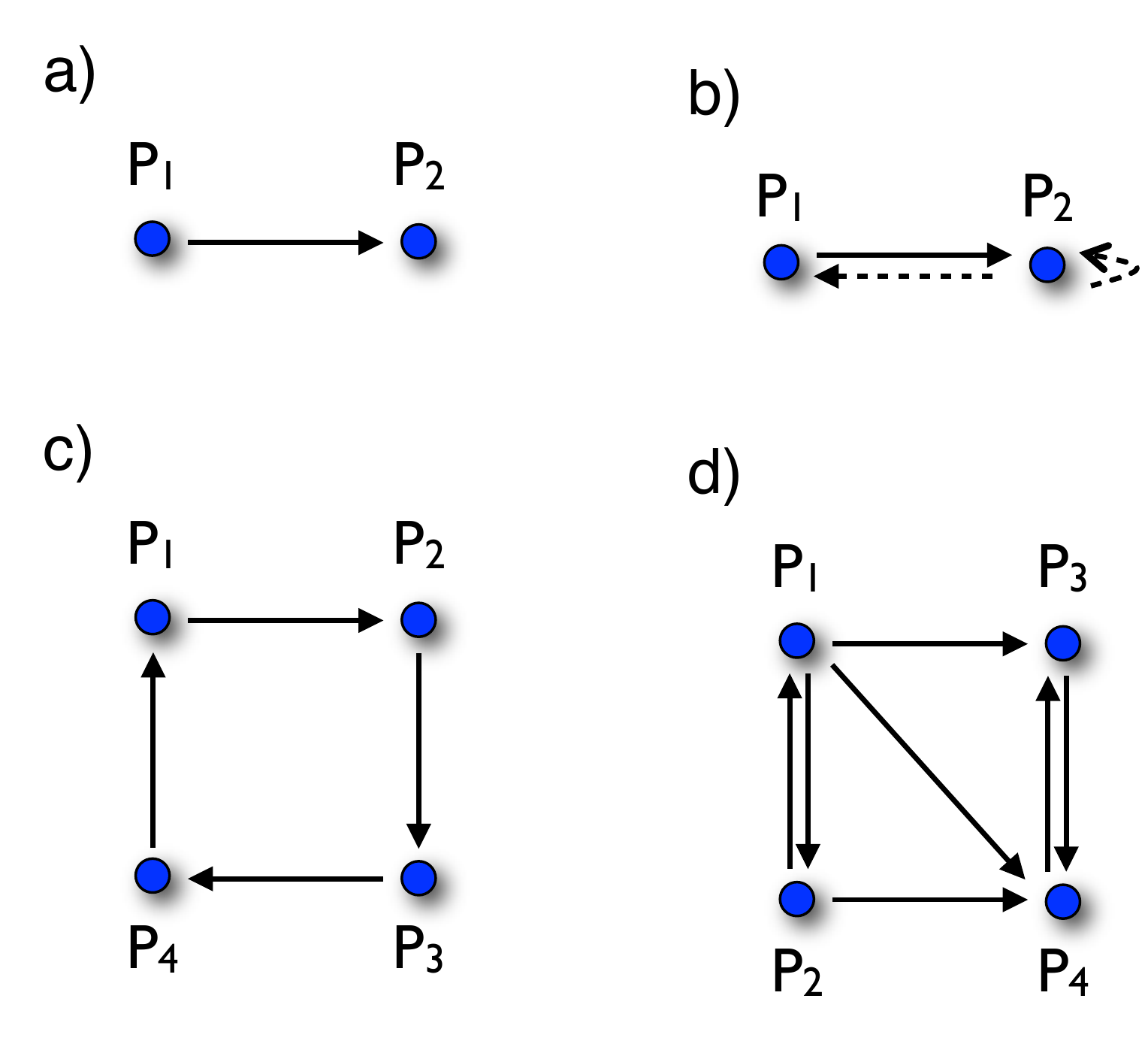}
\caption{(Color online) 
a) A simple web of 2 nodes.
b) The web of 2 nodes with the patch that ensures stochasticity of the matrix $E$.
c) A graph whose matrix $E$'s second eigenvalue, $\lambda_2$, is one (see text in sect.~\ref{sect:one}).
d) A graph that is not \emph{strongly connected}, or equivalently, whose matrix $E$ is \emph{reducible} (see text in sect.~\ref{sect:one}).
Figure adapted from Ref.~\cite{gdpmamd2011}.
}
\label{fig:fig_synoptic_patches} 
\end{figure}

In this section we will briefly go through the main problems that cause the power method to fail to yield a meaningful stationary state, if the ranking is based exclusively on the bare hyperlink matrix $H$.
Indeed, even in the simple case of networks that have {\it  dangling} nodes,
{\it i.e.} pages not linking to other ones (as e.g. in Fig.~\ref{fig:fig_synoptic_patches}a),
the power method will output the null vector. 
Considering the example in Fig.~\ref{fig:fig_synoptic_patches}a,
whose hyperlink matrix is:
\begin{equation}
H = \left ( 
\begin{array}{cc}
\;0\; \;&0\; \\
\;1\; \;&0\; \\
\end{array} \right),
\end {equation} 
if one starts from $I_0 = (1,0)^t$ and iteratively applies the $H$ matrix, the power method will yield the vector $I = (0,0)^t $.
In order to prevent this from happening a strategy is to replace 
the columns corresponding to  dangling nodes with columns whose entries are $1/N$, where $N$ is the number of nodes. 
This is equivalent to having every dangling node linking to every node in the web, including itself. The possibly disconnected graph becomes effectively connected at the price of giving a very low weight to the links added by hand (see Fig.~\ref{fig:fig_synoptic_patches}b). 
The modified hyperlink matrix, $E$, now reads as follows:
\begin{equation}
E = \left ( 
\begin{array}{cc}
0&1/2\\
1&1/2\\
\end{array} \right) \, .
\end {equation} 
It is in general, (column) stochastic, i.e. its columns all sum up to one. It follows~\cite{meyer2000matrix} that the eigenvalue spectrum always contains the value $1$. Furthermore,  the convergence of $I^k =  E I^{k-1}$ to $I$ is dominated by a fundamental quantity, the {\it spectral gap}, {\it i.e.} the difference between $1$ and the absolute value of the second eigenvalue of $E$, $|\lambda_2|$. If the spectral gap is open, {\it i.e.}  $|\lambda_2|$ is strictly smaller than $1$, the power method will converge. In addition, its rate of convergence is larger if the spectral gap is wider, {\it i.e. }$|\lambda_2 |$ is as close to $0$ as possible.  
However, another type of problem can still arise for cyclic graphs, such as the example shown in Fig. 1c, with $E$ matrix:
\begin{equation}
E = \left ( 
\begin{array}{cccc}
0\;&\;0&\;0&\;1\\
1&0&0&0\\
0&1&0&0\\
0&0&1&0\\
\end{array} \right) \, .
\end {equation} 
In this case it is easy to see that, since $|\lambda_2|$ is equal to one, the spectral gap is closed. As a consequence, the power method will not converge as can be easily seen starting from {\it e.g.} $I_0 = (1,0,0,0)^t$.
In order to ensure convergence and that the spectral gap is open, one must require a  property of the matrix $E$, \emph{primitivity} or \emph{regularity}, i.e. that there is an integer $m$ such that $E^m$ contains only positive entries. 

It is helpful to interpret the application of the modified hyperlink matrix as a time step evolution operator of a random walker diffusing on the network. The meaning of this assumption is that the walker will reach any page using a path of at least $m$ links, independently of the starting point.
Let us now consider the graph in Fig.~\ref{fig:fig_synoptic_patches}d.
\noindent It is possible to partition the graph into two subgraphs: $\mathcal G_1$ and $\mathcal G_2$. There are no links pointing from the subgraph $\mathcal G_2$, made of the nodes $3$ and $4$ to the first subgraph $\mathcal G_1$, made of nodes $1$ and $2$. This is evident inspecting the $E$ matrix, indeed, writing down the matrix $E$:
\begin{equation}
E = \left ( 
\begin{array}{cccc}
0\;\;&\;\;1/2\;&\;\bf{0}\;&\;\bf{0}\;\\
1/3&0&\bf{0}&\bf{0}\\
1/3&0&0&1\\
1/3&1/2&1&0\\
\end{array} \right),
\end {equation} 
one can see that the sector of the matrix relative to the connectivity structure of 
the nodes of $\mathcal G_2$ to the nodes of $\mathcal G_1$ is zero, indicating that there are no paths from one subgraph to the other.
This fact has a strong effect on the possibility of interpreting the vector yielded by  
the power method as the vector of the nodes' importances. Indeed, starting from $I_0 = (1,0,0,0)^t$ we find: $I=(0,0,3/5,2/5)^t$. 
This entails that the nodes from the first subgraph $\mathcal G_1$ have zero importance albeit being linked by other nodes. This is clearly unsatisfactory. 
The  \emph{reducibility} of $E$ is the culprit of the {\it drain} of importance from $\mathcal G_1$ to $\mathcal G_2$. 
In order to have a meaningful vector $I$, one that has all nonzero entries, it is necessary that the matrix be \emph{irreducible}. This is equivalent to the fact that the graph is \emph{strongly connected}, i.e. that given two pages there is always a path connecting one to the other  (see~\cite{meyer2000matrix} chap. 8).

\subsubsection*{The Patched Algorithm}
Let us discuss how to implement all the patches and ensure that the power method converges and outputs the PageRank vector, 
whose components are the nodes' rankings.
Let us reuse the random walker's interpretation of the diffusion process on the graph. Let us imagine that the walker
follows the web with stochastic matrix $E$ with probability $\alpha$ and 
that with probability $1-\alpha$, it  will jump to any node at random. The matrix of this process would be:
\begin{equation}
G := \alpha E + \dfrac{1- \alpha}{N} \tilde{\mathbb{1}} ,
\label{ eq: google matrix pagerank}
\end{equation}
 where $ \tilde{\mathbb{1}}$ is the $N \times N$ matrix with all its entries set to $1$. The matrix $G$ is known as the {\it Google Matrix}. It is irreducible by construction because of the addition of the second term in eq.~\eqref{ eq: google matrix pagerank}. For the same reason, it is also primitive or regular, having all positive entries.
By the Perron-Frobenius theorem~\cite{meyer2000matrix} the matrix $G$ has a unique stationary vector $I$ satisfying $G I = I $ and it can be calculated with the power method starting from any nonzero initial condition, $I_0$. 
 
Let us note that the parameter $\alpha$, called the {\it damping parameter}, has been put in by hand and therefore needs to be tuned. The actual value has a strong effect on the PageRank's calculation. Indeed, it is known~\cite{haveliwala2003second} that the spectral gap of $G$ depends on the value of the damping parameter $\alpha$. Indeed, $|\lambda_2|$ is such that $|\lambda_2 | \le \alpha $. In order to speed up the convergence of the power method one would choose $\alpha $ as close to zero as possible but in this way the structure of the web, described by $E$ would not be taken into account at all. The damping parameter $\alpha$ was tuned by hand to the arbitrary value of $\alpha = 0.85$ to optimize the computation of the PageRank vector. 
Loosely speaking, for this value, the random walker performs most of his steps according to the graph-dependent stochastic matrix $E$ and hops to any other node or page of the graph approximately once every seven steps. This ensures that, on the one hand, the actual graph structure still dominates the search and ranking process. On the other, it guarantees that the completely random dynamics according to the second term in Eq.~\eqref{ eq: google matrix pagerank} is large enough to ensure that the walker is able to exhaustively explore the entire graph, independently of possible problems arising from the specific graph and associated hyperlink structure.
Notwithstanding, the effect on the rankings was found to be very dependent on the value chosen for this parameter~\cite{shepelyansky}.

\subsubsection*{Formulation as a Random Walk}

In order to proceed to a valid quantization it will be instrumental to see in detail why Google PageRank is a random walk based algorithm. Indeed, it is useful to identify the importance of a page as the probability of finding a random walker diffusing on the graph according to the stochastic dynamics given by the Google matrix $G$. 
Let us start by defining a set of random variables: $X^{(0)}, X^{(1)}, \ldots, X^{(T)}, \ldots  $, where $T$ is the number of time steps. For each step,  the random variable can take on values in the set of nodes $\{ P_i \}$ of the web. 
We can recast Google PageRank in the language of a Markov Chain. Thus, using the interpretation of the importance of a node given above, one can write the time step evolution of the power method as:
\begin{equation}
{\rm Pr}(X^{(n+1)}=P_i) = \sum_{j} G_{ij} {\rm Pr}(X^{(n)}=P_j)
\label{ eq: pagerank reformulation markov chain}
\end{equation} 
and from the \emph{law of total probability}: 
\begin{equation}
\begin{matrix}
{\rm Pr}(X^{(n+1)}=P_i) = \\
\sum_{j} {\rm Pr}(X^{(n+1)}=
P_i | X^{(n)}=P_j) \, {\rm Pr}(X^{(n)}=P_j),
\label{ eq: pagerank reformulation markov chain 2 }
\end{matrix}
\end{equation} 
one can interpret the stochastic matrix $G$ as the conditional probability linking one time step to the other, i.e.:
\begin{equation}
G_{ij}  = {\rm Pr}(X^{(n+1)}=P_i | X^{(n)}=P_j) 
\label{ eq: pagerank reformulation markov chain G}.
\end{equation}
Furthermore, the PageRank vector is the stationary (or steady state) distribution of the Markov chain.

The latter interpretation of Google PageRank as a random walk is of key importance if we want to find quantum counterparts of the algorithm.

\subsection*{Quantum PageRanks - General Requirements}
\label{sect_II_Quantum_PageRanks}

 

As we have seen, the PageRank algorithm is a random walk based algorithm.
We will discuss our proposal for a quantum counterpart but before delving into the details we will introduce the basic requirements any quantization of the algorithm should satisfy.
Moreover,  these  requirements reflect  
the scenario we will most likely witness in the future.
Indeed, from today's perspective it seems probable that quantum networks of considerable sizes will be operational on a near-future time scale, before fully-fledged large-scale quantum computers might become available, whose construction despite enormous progress still poses considerable technological challenges. 
These properties we require are meant as  a guide to introduce a realistic class of Quantum PageRanks and read as follows:

{\bf Quantum PageRank Class}:
\noindent \begin{description}

\item[P1] The classical PageRank must be embedded into the quantum class
in such a way that the directed graph structure is preserved at the quantum level.

\item[P2] The sum of all quantum PageRanks must add to 1, {\it i.e.} $\sum_i I_q(P_i) = 1$.

\item[P3] The Quantum PageRank admits  a quantized Markov Chain (MC) description. 

\item [P4] The classical algorithm to compute the quantum PageRank belongs to the computational complexity 
class P. 

\end{description}

Property P1 reflects the fact that originally we would have an internet that is a classical network represented by a 
directed graph and then shall apply a certain quantization procedure in order to turn it into a quantum network.
The latter must be compatible with the classical one, particularly preserving the directed structure which is crucial to measure a page's  authority. 
Moreover, as we have seen, the random walk interpretation naturally entails that the hopping probability from one node to the other may not be symmetric. Therefore, if we want to define a quantum algorithm to analyze also directed classical networks it is very important to preserve the directionality of the links. Preserving the directnedess is a nontrivial task
and some quantization methods may fail to produce a unitary quantum PageRank importance for the quantum case. 
An example is the quantum walk on a line whose naive quantization fails to yield a unitary time evolution operator, see~\cite{gdpmamd2011} for details.

With property P2 we guarantee that we have a globally well-defined notion of the importance of a web page at the quantum level.
This allows us to have the probabilistic interpretation of the  walker's position which is essential in PageRank's definition of the importance of a node (see Sect.~\ref{sect:one}).

Property P3 is the key to a wide class of natural quantization methods for the classical PageRank based on the equivalence of this one
with a classical Markov chain process (see Sect.~\ref{sect:one}). Thus, it is natural that the equivalent property holds true in the quantum version of the PageRank,
and consequently, its description in terms of a quantized walker's motion.

The reason for requiring property P4 relies on the assumption that we envisage a near-future scenario when a certain class of quantum network will be operative
but not yet a scalable quantum computer. Therefore, we demand that the computation of the quantum PageRank $I_q$ can be efficiently carried out
on a classical computer.

In~\cite{gdpmamd2011} we have constructed a valid quantum PageRank that fulfills all these requirements. We remark that there may be other solutions to the problem of finding a  quantum version
of PageRank within the class defined above, but nevertheless finding one instance of this quantum PageRank class is already a nontrivial task.

In the following  we will
describe in detail a  way to devise a quantum algorithm by making use of Markov Chain quantization methods. In particular, we will use results introduced by Szegedy in the context of quantum walk based algorithms for the detection problem to find a valid quantization of Google's  PageRank.  We will then review
the main features we have found when applying the algorithm to networks of various sizes and topologies. 

\subsection*{A Quantization of Google PageRank}
\label{sect_II_A_Quantization_of_Google_PageRank} 

In the following we define a valid Quantum PageRank algorithm using the quantum walk formalism introduced by Szegedy~\cite{2004_Szegedy_IEEE}. Furthermore, we hint at the analysis of the walk's spectrum that is instrumental in providing us with an efficient algorithm.

A natural way to define a quantum counterpart  of the importance of a node or page in the quantum network associated to a directed graph
is to exploit the connection with the Markov chain process in which the  walker can now take advantage of quantum dynamics over the nodes of the network.
Similarly to many quantum algorithms, we will start from an initial vector $ \, | \psi_0 \rangle$ let it evolve according to an evolution operator. We will then project it onto a state encoding the node and we will interpret the resulting probability distribution as the {\it instantaneous Quantum PageRank} of the node, which will be a new measure of its importance.

Let us start by introducing the key ingredients to quantize the Markov Chain \`a la Szegedy. Let $G$ be an $N \times N$ stochastic matrix representing a Markov chain, in our case the Google matrix, on a network of $N$ nodes.
Let the Hilbert space be the span of all vectors representing the (directed) edges  of the graphs i.e. ${ \mathcal H }=  {\rm span}\{ | i\rangle_1 | j \rangle_2 \, ,{\rm with}\, i , j \in N\times N\} = {\mathbb C}^{N} \otimes {\mathbb C}^{N}$. This definition of the Hilbert space allows us to preserve the directedness of the walk as we will see. To emphasize it we will explicitly write the subindices referring to the spaces in the tensor product. 
 Let us define the vectors
\begin{equation}
| \psi_j \rangle :=  |  j \rangle_1 \otimes  \sum_{k=1}^{N}  \sqrt{ G_{kj} } \,| k \rangle_2 .
\label{ eq: Sgezedy vector 1}
\end{equation} 
Here $| \psi_j \rangle$ is a superposition of the vectors representing the edges outgoing from the $j^{th}$ vertex. The weights are given by the (square root of the entries of the) Google matrix $G$.

One can easily verify that due to the stochasticity of $G$ the vectors $ | \psi_j \rangle \,$ for $ j=1,2,\ldots , N$ are normalized and form an $N-$dimensional orthonormal set of vectors. 
The initial state of our walk is:
 \begin{equation}
| \psi_0 \rangle = \sum_j | \psi_j \rangle \, .
\label{eq:initial_condition_szegedy}
\end{equation} 
Let us now define the quantum walk operator that governs the evolution. First, let us first introduce the operator 
\begin{equation}
\Pi := \sum_{j=1 }^{N} | \psi_j \rangle  \langle  \psi_j |,
\end{equation}
which  is  a projector onto the subspace generated by the  vectors $ | \psi_j \rangle$ for $ j=1,2,\ldots , N$. The single-step evolution operator of the quantum walk is given by 
\begin{equation} 
U := S (2 \Pi -\mathbb{1})
\label{ eq: Sgezedy unitary walk operator}
\end{equation} 
where $S$ is the swap operator, {\it i.e.} 
\begin{equation} 
S=\sum_{j,k=1}^{N} | j, k \rangle \langle k, j |.
\end{equation} 
The time step is thus the effect of a reflection followed by a swap operator. Let us look more closely at the reflection operation: 
\begin{equation}
2 \Pi -\mathbb{1} = \sum_{j=1 }^{N} \left( 2\, | \psi_j \rangle  \langle  \psi_j | - \dfrac{1}{N} \mathbb 1 \right).
\label{ eq: Sgezedy coin flip}
\end{equation} 
The vectors $| \psi_j \rangle  $ contain the information of the directed links that connect the $j^{th}$ node to all its neighbors to which it is connected through the stochastic matrix $G$. The sum over all nodes of the operators $ 2\, | \psi_j \rangle  \langle  \psi_j | - \frac{1}{N} \mathbb 1 $ is nothing but a reflection around the subspace spanned by the vectors $| \psi_j \rangle  $ and has the effect of enhancing the amplitudes of the mentioned directed edges at the expenses of the others.

We are now in a position to outline the whole algorithm to compute the instantaneous Quantum PageRank of a node.
In order to implement the  procedure, one starts from the stochastic matrix $G$ encoding the network and the Google walk that we want to quantize. The initial state of the walk is  $ \, | \psi_0 \rangle$ and its evolution is performed according to the two-step evolution operator $U^2$. 
The evolution according to a 
two-step evolution operator 
is an essential property of the algorithm and its associated quantum walk dynamics: This characteristics guarantees that the directions of the edges of the graph are swapped an even number of times, which in turn preserves the directedness of the graph. 
The instantaneous PageRank of  the node $i$ is given by the probability of finding
the walker on  the node $i$ of the network after $m$ time steps. In order to compute this probability 
we need  to project onto $| i \rangle_2  $, and  finally to take formally the squared norm of the resulting quantum state:
\begin{equation}
I_q(i,m) =  \langle  \psi_0 | \,  {U^\dagger}^{2m} | i \rangle_2  \langle  i |  U^{2m} | \psi_0 \rangle .
\label{ eq: Quantization of PR importance}
\end{equation}
The dynamics is unitary and therefore
we expect its instantaneous value to oscillate in time.
We also note that it is possible to give an equivalent way of expressing the instantaneous Quantum PageRank as follows:
\begin{equation}
I_q(i,m) =  \mathrm{Tr}_1 \,  \left( U^{2m} | \psi_0 \rangle  \langle  \psi_0 | {U^\dagger}^{2m}  | i \rangle_2  \langle  i | \right) .
\label{ eq: Quantization of PR importance_alt}
\end{equation}

It is possible to enforce a probabilistic interpretation of the Quantum PageRank as required by the property P2. Indeed,
\begin{equation}
\sum_i I_q(i,m)  = \langle  \psi_0 | \,  {U^\dagger}^{2m} | \sum_i |  i \rangle_2  \langle  i |  U^{2m} | \psi_0 \rangle =  1 \quad \forall m .
\label{ eq: Quantization of PR importance sum check}
\end{equation}
This allows us to interpret the quantity $I_q(i,m)$ as the instantaneous relative importance of the node  $i$, identifying it with the probability of finding a quantum walker on node $i$.

In order to integrate out the fluctuations arising from the coherent evolution we also introduce the \emph{average} importance of the $i^{th}$ node  $\langle I_q(i) \rangle$ as:
\begin{equation}
\langle I_q (i)\rangle:=\frac{1}{M} \sum_{m=0}^{M-1}  \ I_q (i,m).
\label{mean_value}
\end{equation}
We also note that the average Quantum PageRank converges to its asymptotic value as $M$ grows larger~\cite{Aharonov2002}.

We note that it is possible to identify an invariant subspace, $\mathcal{H}_{dyn}$, where the dynamics of the quantum walk takes place. Indeed, the initial state of the walker lies within this invariant subspace. Therefore, it is possible to calculate the instantaneous Quantum PageRanks  using the eigendecomposition of the  operator $U^2$ restricted to $\mathcal{H}_{dyn}$ which has at most dimension $2 N$, where $N$ is as usual the number of nodes. For the details on the construction of $\mathcal{H}_{dyn}$ see Ref.~\cite{gdpmamd2011}.
The existence of this subspace greatly simplifies the numerical task of simulating the Quantum PageRank on a classical computer. Indeed, it was of key importance in extending the results to bigger complex networks~\cite{paparo2013quantum} 
since this allows to work with matrices with dimension - at most - $2N$ as opposed to $N^2$.

\noindent Let us state the whole procedure to calculate the instantaneous Quantum PageRank and its average.

\noindent {\em \bf Quantum PageRank Protocol}
\begin{description}
\item[Step 1/]  Write the Google matrix of the network ${\cal G}$.
\item[Step 2/] Write down the initial state $ | \psi_0 \rangle$ and the evolution operator $U^2$.
\item[Step 3/]  Find the eigenvectors and eigenvalues of the two-step quantum diffusion operator $U^2$ in the dynamical subspace $H_{dyn}$
(see Ref.~\cite{gdpmamd2011} for the details of the construction).
\item[Step 4/] Extract the Quantum PageRank value in time \eqref{ eq: Quantization of PR importance} and its mean \eqref{mean_value}.\end{description}

\subsection*{Results: Quantum PageRanks on Small Networks}
\label{sect_II_Results} 

\begin{figure}
\centering
\includegraphics[keepaspectratio=true,width=.99\linewidth]
{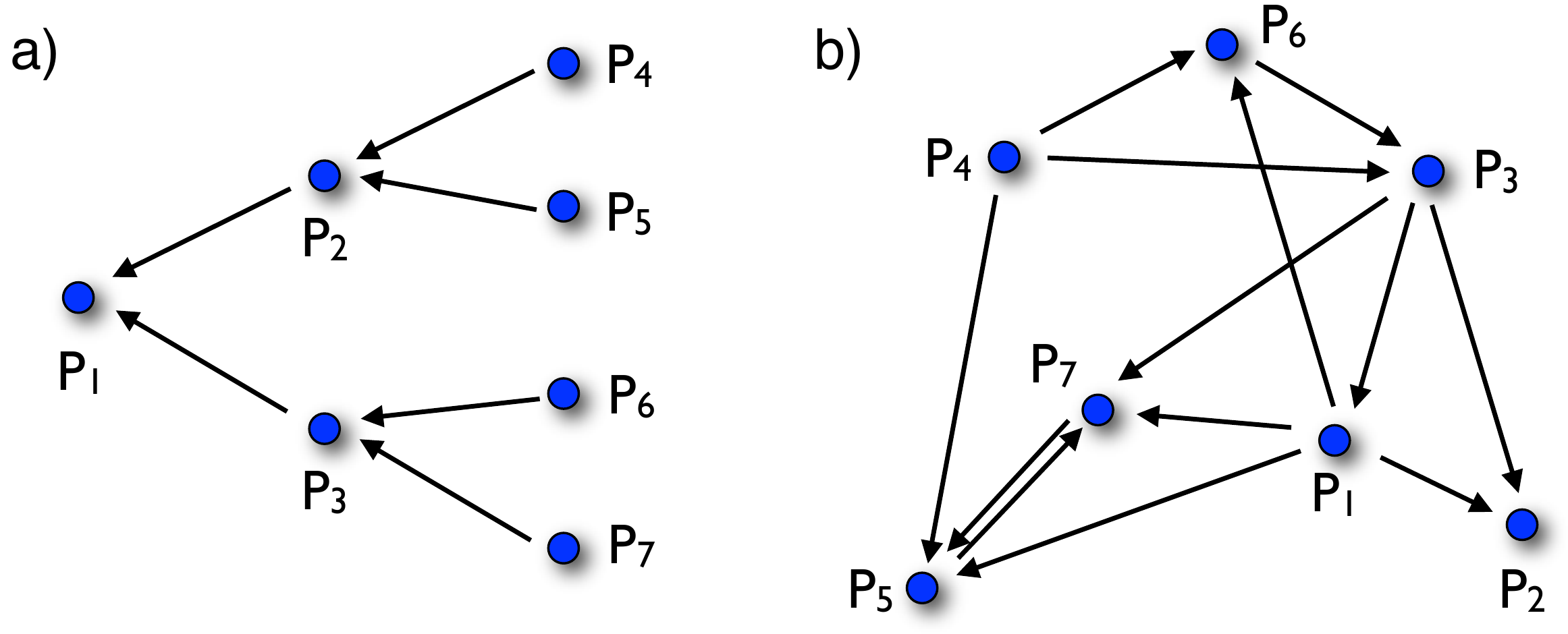}
\caption{
(Color online) Small-scale directed networks used for benchmarking of the Quantum PageRank algorithm, see also discussion in Sect.~\ref{sect:one}. a) The three level binary tree, where each node represents a web page in an intranet with the root node being its home page.
b) A small-size  network not displaying any particular symmetry.
Figure adapted from Ref.~\cite{gdpmamd2011}.
}
\label{fig:fig_7node_graphs}
\end{figure}

After developing a quantum version of  Google PageRank, it is necessary  to apply it to specific networks by means of simulations and to
see how it behaves as compared with the classical PageRank algorithm.
Let us start by discussing the results obtained in~\cite{gdpmamd2011} of our new quantum version of the PageRank algorithm in the case of a binary directed tree with 3 levels (see Fig.~\ref{fig:fig_7node_graphs}a) and of a small directed graph displaying no symmetry (see Fig.~\ref{fig:fig_7node_graphs}b). 

The tree graph (see Fig.~\ref{fig:fig_7node_graphs}a) has a clear meaning in terms of a web network: it represents an intranet with the root node being the home page of a 
website and its leaves representing internal web pages. This case has been extensively studied classically \cite{marijuan11} to devise strategies to improve the root's importance.
We analysed the network using our quantum algorithm. The numerical simulation showed~\cite{gdpmamd2011} that the quantum PageRank of the root page clearly oscillates in time and attains values that are higher than the classical counterpart (see  Ref.~\cite{gdpmamd2011} for details). This property of the quantum PageRank, called {\it  instantaneous outperformance} can be exploited to identify the root by anomalously high values of its importance, measured by quantum means. 

\begin{figure}
\centering
\includegraphics[keepaspectratio=true,width=.99\linewidth]
{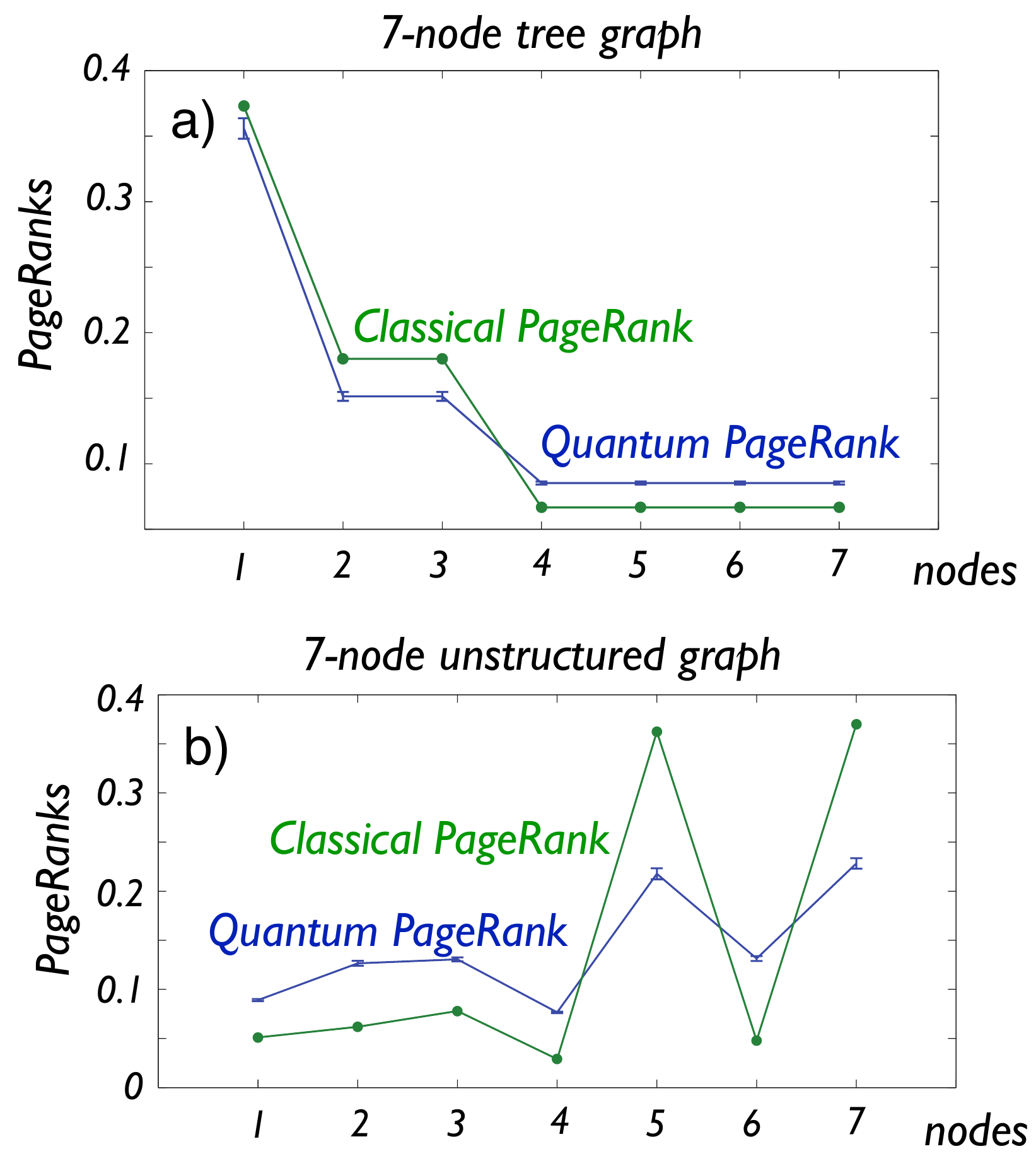}
\caption{
(Color online) The average Quantum PageRanks of the networks in Fig.~\ref{fig:fig_7node_graphs} used to benchmark the Quantum PageRank algorithm (see text in sect.~\ref{sect:one}). a) The three level binary tree.
b) A small-size  network not displaying any particular symmetry.
Figure adapted from Ref.~\cite{gdpmamd2011}.
}
\label{fig:figure_7node_graph_results}
\end{figure}

A distinctive feature of the Quantum PageRank that was highlighted in~\cite{gdpmamd2011} is that the hierarchy is not preserved at all times as measured by the instantaneous value. However,  the hierarchy of the nodes was found to be preserved when measured by the average Quantum PageRank. Indeed, the hierarchical structure was found to be clearly visible (see Fig.~\ref{fig:figure_7node_graph_results}).

In~\cite{gdpmamd2011} a computation of the quantum PageRank was performed also in the case of a general directed graph with no particular symmetry (see fig.~\ref{fig:fig_7node_graphs}b). 
Also in this case the property of {\emph instantaneous outperformance} was found. Indeed, the node with the highest classical PageRank attains, at given times, values of the Quantum PageRank that are higher than the classical counterpart.

\noindent The classical hierarchy is not preserved by the QPR at any given time. Indeed, it was found that the Quantum PageRank predicted a different rankings' list. Moreover, the overall importance was found to be more homogeneously  distributed among all the nodes. Contrary to the classical algorithm that displays the tendency to concentrate a large fraction of the total importance on the nodes with a high ranking, the quantum one was shown to distribute it more evenly, thereby achieving the same task of ranking using a smaller range of values of importance (see Fig.~\ref{fig:figure_7node_graph_results}). This feature was found to be at the base of the increased visibility of the networks' structure in graphs of scale-free topology, for example, or its capability to lift the degeneracy of nodes. These features motivated the extension of the study to bigger complex networks~\cite{paparo2013quantum}.


\section{Quantum PageRank on Complex Networks}
\label{sect:3}


Motivated by the results on small-size networks, in Ref.~\cite{paparo2013quantum} we analysed complex networks using the Quantum PageRank algorithm. In this section we will review the main findings focusing on 
random scale-free networks, because of their widespread appearance and relevance in real-world applications, and on hierarchical networks.
The latter case is an interesting case {\it per se} and the study constitutes a generalization of the one, reviewed above, performed on the binary tree graph.

\subsection{Scale-free Networks}
\label{sec:3B}

\begin{figure*}
\includegraphics[keepaspectratio=true,width=0.99\linewidth]{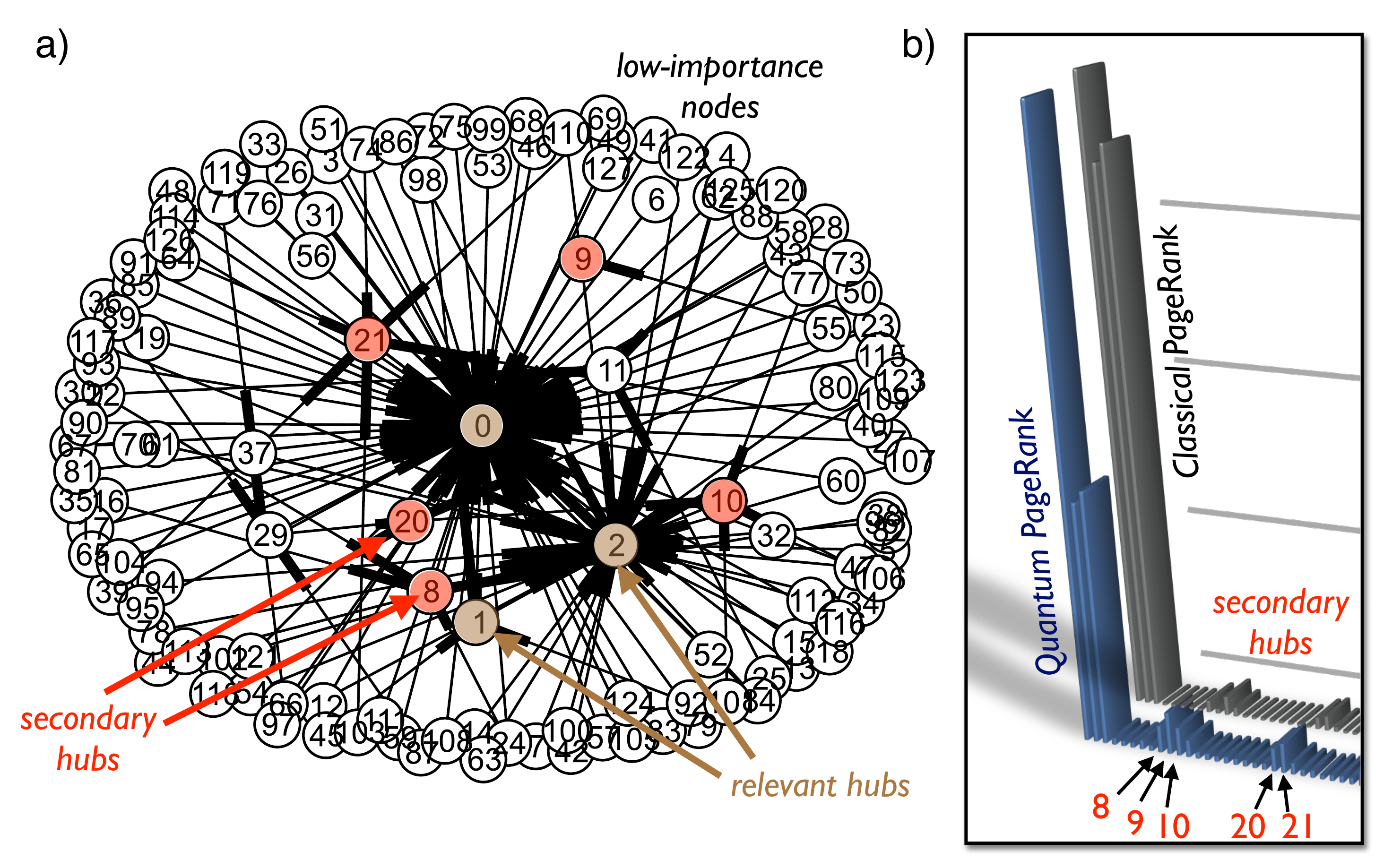}
\caption{
\label{fig:SF_128_nodes_Class_PR_vs_QPR}
(Colour online) a) Scale-free graph with 128 nodes (see text in the sect. \ref{sect:3}) and b) a comparison of the importance of the nodes when evaluated with the quantum and classical PageRank. The classical PageRank shows a very sharp concentration of importance on the three nodes 0, 1 and 2 (relevant hubs). One can see from the comparison of the predictions of the two algorithms the relative emergence of secondary hubs (nodes 8, 9, 10, 20 and 21) when the importance is calculated with the quantum PageRank (see text in the sect. \ref{sect:3}). Figure adapted from Ref.~\cite{paparo2013quantum}.
}
\end{figure*}
\begin{figure*}
\includegraphics[keepaspectratio=true,width=.99\linewidth]{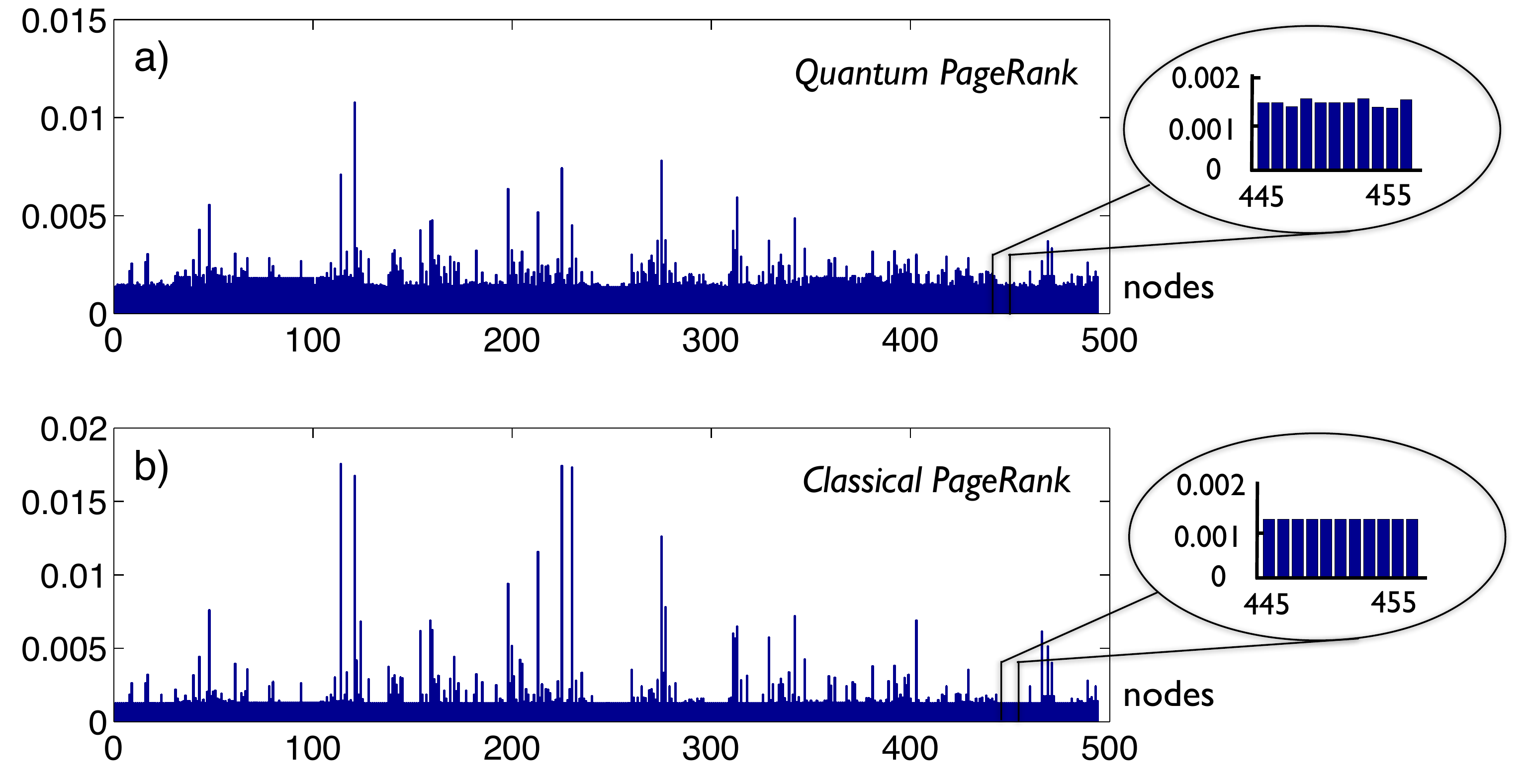}
\caption{
\label{fig:MM_EPA_900_2300_Class_PR_vs_QPR}
(Color online) Comparison of the a) quantum and b) classical PageRank on a real network originating from the hyperlink structure of www.epa.gov~\cite{Batagelj_Mrvar_2006}. One can clearly see how the hubs in the classical algorithm tend to concentrate nearly all the importance. The insets show that the quantum algorithm is capable to lift the degeneracy of nodes in the the low part of the list~(see text in the sect.~\ref{sect:3})
Figure adapted from Ref.~\cite{paparo2013quantum}.
}
\end{figure*}

Random scale-free graphs~\cite{Barabasi_RMP_2002,Boccaletti_2006} are networks that display a small fraction of hubs, i.e. nodes with a high connectivity. This fact is reflected 
by the scale-free behavior of the degree distribution $P(k)$, 
{\it i.e.} $P(k) \approx k^{-\gamma }$. 
These graphs have stimulated a lot of research lately since they seem to be found ubiquitously in nature. Interestingly, the World Wide Web belongs to this class~\cite{Barabasi_Albert_Jeong_2000}. However, examples of scale-free networks are not restricted to the internet and include airline~\cite{Barrat_2004} or metabolic networks~\cite{Jeong_Mason_2001,Jeong_Tombor_2000}, just to name a few. 
The networks in this class display remarkable properties, such as robustness against uncoordinated attacks~\cite{Albert_Jeong_Barabasi_2000,Callaway_Newman_2000,Vazquez_Moreno_2003}, good navigability \cite{Boguna_2009,Carmi_Carter_2009,Lee_Holme_2000} and controllability \cite{Liu_Slotine_2000,Nepusz_2012,Nicosia_Latora_2012}.
To reproduce the characteristic  power-law distribution of the degrees of nodes, various models have been proposed, such as the \emph{preferential attachment model}~\cite{Barabasi_Albert_99,Barabasi_13}. 
In this case, during the network's growth, links are preferentially added to  nodes with a high connectivity.
A random directed scale-free model for the WWW was also introduced in~\cite{diameterWWW} and a generalisation 

appeared in~\cite{BoBoChRi03}. 
In Ref.~\cite{paparo2013quantum}  we simulated the  classical and  Quantum PageRank algorithms on random directed scale-free networks, with sizes ranging up to hundreds of nodes, which  were generated  and analysed using the NetworkX package~\cite{HaScSw08}.
It was clearly shown that the algorithm was able to correctly uncover the scale-free topology of the network  
identifying the most important hubs. 
As opposed to the classical PageRank, however, the quantum PageRank algorithm is capable of unveiling the structure of the graph to a finer degree, highlighting also the relevance of secondary hubs (see Fig.~\ref{fig:SF_128_nodes_Class_PR_vs_QPR}).
The other main finding was that the quantum algorithm is capable of lifting the degeneracy of the nodes that have a lower importance, as can be seen in Fig.~\ref{fig:MM_EPA_900_2300_Class_PR_vs_QPR}.
These results were obtained analysing a subgraph of the WWW obtained by exploring pages linking to www.epa.gov (datasets available from Pajek~\cite{Batagelj_Mrvar_2006}).


\subsection{Hierarchical Networks}
\label{sec:3C}

Many real-life networks are  scale-free but have also a modular structure~\cite{Ravasz_2002,Song_2005}. 
Examples range from technological networks, such as the WWW and electronic circuits, to biological systems like protein and metabolic networks. 
These networks are composed of modules that group different sets of nodes. Moreover, these modules can be distinguished  by the property that nodes belonging to the same module are usually strongly connected. On the other side, modules are in general weakly connected among each other. Therefore, even when the networks are scale-free, their hubs tend to have a low clustering as they joint different modules. Some authors claim that a signature for a hierarchical network is that, apart from displaying the small-world and scale-free properties,  the scaling of the clustering of the $i^{th}$ node $C_i$ with its degree $k_i$ follows $C_i \propto 1/k_i$~\cite{RaBa03,Barabasi-Oltvai-04}.

Hierarchical network models are usually amenable of analytical treatments because they are constructed using  recursive rules. 
For example, we can start from a complete graph $K_n$  
and connect $n-1$ replicas of $K_n$ to a selected root node.
Next, $n-1$ replicas of the new whole structure can be added to this root. 
At this step the  graph will have $n^3$ vertices. The process goes on until we reach the desired graph order. 
One can define many different types of  hierarchical networks, depending, {\it e.g.} on the initial graph used, the introduction of extra edges  among the different copies of the complete subgraphs, etc. However, once the starting graph is given, these networks do not display other free parameters that can be  adjusted and their main characteristics are fixed.

\begin{figure*}
\centering
\includegraphics[keepaspectratio=true,width=.99\linewidth]
{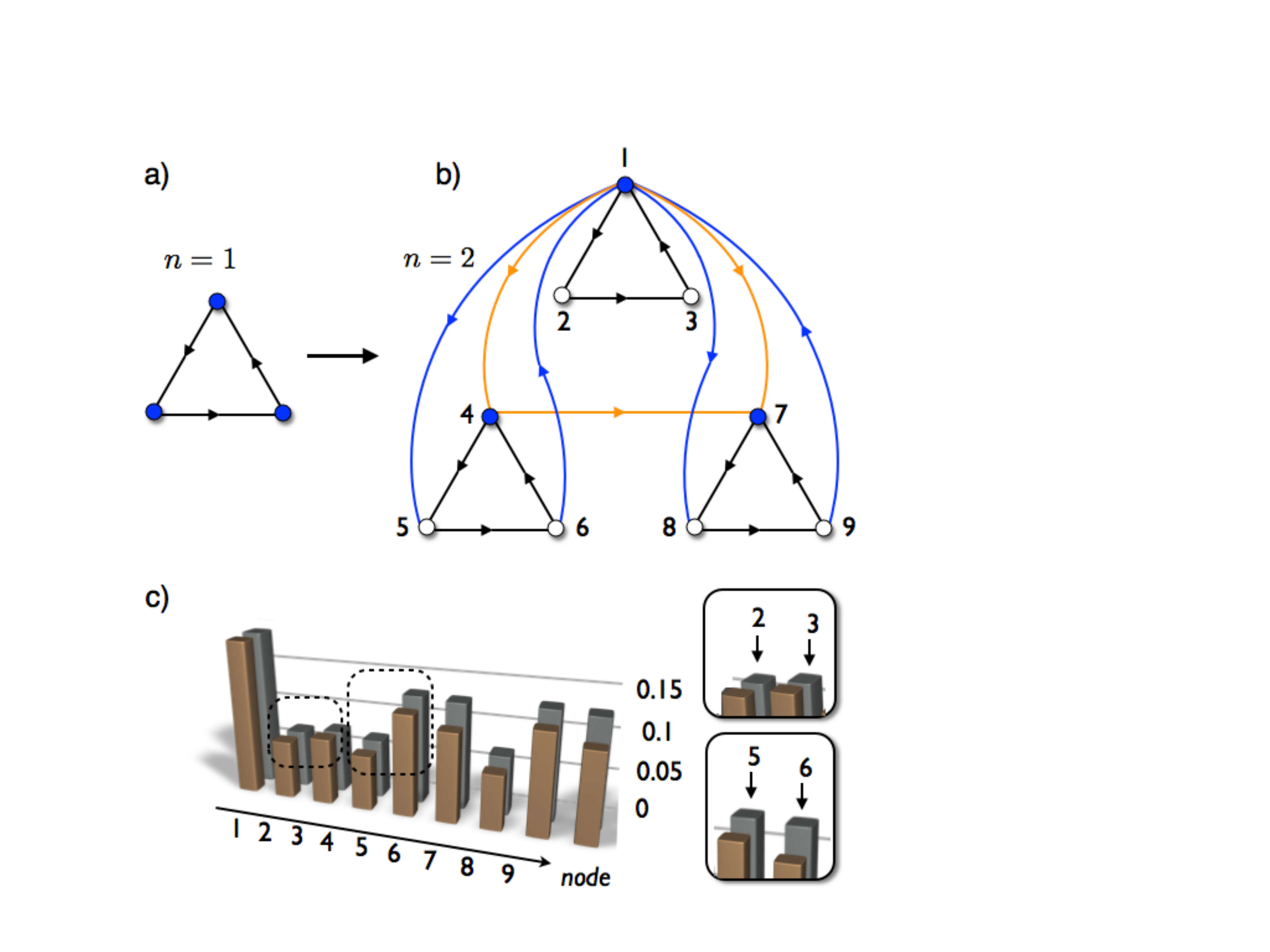}
\caption{
(Colour online) Construction of  hierarchical networks. 
The family of directed hierarchical graphs. In this case the generation labeled by $n$ has $3^{n}$ nodes. We consider graphs of the generations with $n=2,3,4$.
Figure adapted from Ref.~\cite{paparo2013quantum}.
}
\label{fig:HierarchicalConstruction2}
\end{figure*}

In \cite{BaRaVi01}, Barab\'asi et al.  introduced a  family of 
hierarchical networks and proved it had a small-world scale-free nature. The model was generalised in~\cite{RaBa03} and its study extended in~\cite{No03}.  For our analysis we have designed a directed version based on these graphs, see figure~\ref{fig:HierarchicalConstruction2}. In this case the starting point is a directed 3-cycle.

Another remarkable family of hierarchical directed graphs has been obtained by considering the directed version of  the construction published in~\cite{Comellas_Miralles_PhysA,Comellas_Miralles_JPhys}.
In this case, the graphs are small-world, self-similar, unclustered and outerplanar, {\it i.e.}  there exists an embedding where all vertices lie on the boundary of the exterior face. However, these graphs  are not scale-free, but follow an exponential distribution.

In Ref.~\cite{paparo2013quantum} we extended the study to hierarchical networks using the quantum PageRank algorithm. We focused on two families of graphs (one of which is represented in figure~\ref{fig:HierarchicalConstruction2}a and b) and it was shown that the structure of hierarchical moduli  is clearly displayed by the average PageRanks and that the structure is preserved. Remarkably, at the local {\it intra-layer} level, the quantum PageRank is able to highlight the connectivity structure of the nodes that belong to the same level in the hierarchical construction (see figure~\ref{fig:HierarchicalConstruction2}c). 

The quantum PageRanks displayed an amplification in the difference in importance between nodes belonging to the same hierarchical layer but with different local connectivity, thus resulting in an increased resolution of the network structure. 


\section{Further Aspects}
\label{sect:Further_Aspects}
In Ref.~\cite{paparo2013quantum} several other remarkable features of the quantum algorithm were uncovered when the study was extended to networks of larger sizes and different topologies.
In this section we review the main findings regarding the localisation properties of the walker on scale-free networks, the algorithm's stability with respect to the variation of the damping parameter $\alpha$, the scaling behavior of the rankings and the sensitivity to coordinated attacks. In~\cite{paparo2013quantum} these features are  thoroughly discussed and we refer the interested reader to this paper where she or he will find more details on how these results were obtained.

\subsection{Localisation}

The classical and quantum algorithms are based on random -or quantum- walks that take place on a network. This sole fact 
allows to look at the problem from a different angle, making use of tools normally employed in the study of random or quantum walks on physical networks. On the other side, scale-free networks are very prominent in nature and exhibit the property that  few nodes, the hubs, account for the majority of the importance. 
Since the importance can be seen  as the probability of occupation of a node by a random -or quantum- walker diffusing on the network, a necessary condition for any algorithm to rank satisfactorily well is that the walker be localised. Equivalently,
the number of nodes with a non-negligible probability to find the walker should be very small comparing it to the number of nodes in the network. 

In the classical case the localisation-delocalisation transition was studied~\cite{shepelyansky} using a 
tool which is regularly used in condensed matter physics, 
the Inverse Participation Ratio (IPR). This quantity is a function of the probability distribution of the walker on the network and its scaling behavior with respect to the network size is a good witness of the localisation or delocalisation of the walker. In the case of scale-free networks it was shown~\cite{shepelyansky}  that for $\alpha $ ranging from $0.4$ to $1$ the random walker is in a localised phase, thus reinforcing the idea that PageRank's output is a meaningful measure of importance on this type of networks.
In Ref.~\cite{paparo2013quantum} we extended the definition of the IPR, making use of the average Quantum PageRank as a probability distribution, and used it to study the diffusion of the walker on networks of different topologies. In the case of scale-free graphs we used a value for the damping parameter of $\alpha =0.85$ and it was shown that  scale-free graphs favour a localisation phase also in the case that it is a quantum -and not classical- walk taking place on a network with this topology. 
This is consistent with the fact that the quantum algorithm can rank satisfactorily well the nodes of such a network. Indeed, in order to unveil the main hubs the random or quantum walk must be able to localise the walker on few important nodes.

\subsection{Stability with respect to the damping parameter}

In the classical PageRank algorithm the value of the damping parameter $\alpha = 0.85$ was set
to mimic  the behaviour of a walker (or surfer) that randomly hops to any node roughly once every seven times, ensuring the convergence to a meaningful probability distribution. Only \textit{a posteriori} it turned out that this choice was indeed a sensible one given that the WWW network was found to have the small-world property~\cite{diameterWWW}.
Due to the arbitrariness of the choice, though, it is  desirable that the output of the algorithm varies slowly with respect to the variation of the damping parameter. In the classical PageRank case it was found that the effect of this parameter on ranking is large and that two rankings obtained using different values of this parameter can be very different~\cite{shepelyansky}.

In Ref.~\cite{paparo2013quantum} we  tackled this problem in the quantum case making use of  two quantities, related to the  \textit{classical} and \textit{quantum fidelity}~\cite{rmp,NC,lo1998introduction}, the latter being equivalent to the \textit{trace distance}, to measure to which extent the outputs of the quantum algorithm were different when $\alpha $ was varied. 

Our study was performed varying  the value of $\alpha$  from $0.01 $ to $0.98$ for scale-free networks of $128$ nodes and the results  clearly showed that the quantum PageRanks vary less than the classical counterpart. Furthermore, it was shown numerically  that the minimum fidelity between two rankings obtained with any two values of $\alpha$ and $\alpha^\prime$ was always above the value $0.91$ enforcing the idea that  ranking is less dependent on the damping parameter when it is performed using Quantum PageRank.
Similarly, it was investigated to which extent the rankings vary when a different value of $\alpha \neq 0.85$ is chosen. This study was used to compare the quantum algorithm to its classical counterpart and also in this case the quantum algorithm was shown to be less prone to the variation of the damping parameter concluding that the Quantum PageRank  is more robust with respect to variation of the damping parameter $\alpha$ that controls the fraction of random hopping.

\subsection{Scaling Behaviour}

For the classical algorithm, it was shown~\cite{donato2004,Pandurangan2005} that for real networks the nodes' classical PageRanks $I_j$, sorted in descending order,  follow a scale-free or power law  behavior. This is indicative of the fact that the algorithm is able to identify the hubs of a scale-free network. Furthermore, the scaling exponent $\beta$ is a measure of the homogeneity with which the importance is distributed among the nodes of the network. In Ref.~\cite{paparo2013quantum} a similar study was done for the case of the quantum PageRank.

Here we review the main findings obtained performing the analysis on scale-free networks of 256 nodes. The  scaling behaviour of the Quantum PageRanks versus the index  of the node sorted in descending order was found by means of statistical regression. 
The analysis showed that on scale-free networks, Quantum PageRanks displayed a power law behaviour. Furthermore, it was found that the scaling coefficients are different in the quantum and classical case. We showed  that the coefficient for the quantum case is smaller pointing to the fact that the quantum PageRank has a smoother behaviour, giving less relative importance to the nodes in the high part of the ranking list. On the other hand, it was found to be able to better rank in the low part of the list (where the classical PageRank gives highly degenerate values) lifting the degeneracy (see Ref.~\cite{paparo2013quantum} for details).
  
Finally, by considering real-world networks, such as a subgraph of the WWW obtained by exploring pages linking to www.epa.gov~\cite{Batagelj_Mrvar_2006}, it was found that also in this case a power law behaviour was present. Furthermore, the scaling coefficient was found to be smaller in the quantum case enforcing the idea that it is a feature of the algorithm that is present in every network, ranging from numerically generated to real-world examples.

\subsection{Sensitivity  of the Rankings under Coordinated Attacks}


In Ref.~\cite{paparo2013quantum} we have also studied and partially addressed the question of how sensitive the quantum PageRank protocol is with respect to attacks on specific nodes of the graphs. More precisely, we asked to which extent the ranking results provided by the quantum Google algorithm change as a whole, when certain nodes of the graphs are attacked and fail, and the algorithm is subsequently run on the reduced graph formed by the remaining, unaffected nodes. For our analysis we focused on ensembles of scale-free, directed graphs of mesoscopic size, containing up to 32 nodes or pages. The scenario we considered was the one of coordinated instead of random attacks, i.e. the case where the hubs as the most important nodes are attacked and consequently removed from the graph. For this situation we analyzed how the order in importance of the remaining nodes, as provided by both the classical and the quantum PageRank algorithm differed from the rankings provided when applying the classical and quantum algorithms to the original, complete graphs. Our study revealed that the quantum PageRank showed an increased sensitivity of the ranking outcomes under coordinated attacks, as compared to the classical algorithm applied under the same conditions. We conjectured that the observed enhanced sensitivity might be related to the advantageous property of the quantum protocol of being able to reveal more sub-structure of graphs than the classical protocol. Specifically, as discussed above, the quantum ranking algorithm is able to lift the degeneracy in importance of the nodes of low importance values -- see insets in Fig.~\ref{fig:MM_EPA_900_2300_Class_PR_vs_QPR} -- whereas the classical algorithm is known not to resolve such degeneracies to the same extent. As discussed in more detail in Ref.~\cite{paparo2013quantum}, this suggests that attacks of the hubs have a stronger effect in the quantum case, as they manifest themselves as structural changes of the network structure, which in turn can lead to a reordering of the low-importance nodes with only slightly differing importance values.

\section{Conclusion}
\label{sect:4}

In this paper we have reviewed the behaviour of the quantum PageRank algorithm, developed in Ref.~\cite{gdpmamd2011}, and applied to complex networks in Ref.~\cite{paparo2013quantum}. 

Already on small networks the algorithm was shown to display remarkable properties such as \emph{instantaneous outperformance}, i.e. some Quantum PageRanks would attain instantaneous values that were anomalously high. 
In directed trees, which are good models of intranets, this property can be exploited to recognize the most important node, the root. Indeed, a very high value of the instantaneous QPR is a good witness that the node is the root of the tree network.
The algorithm, when applied to general graphs, has also been shown to rank using a more homogenous distribution of importance. These results, among others, motivated the study of the algorithm on bigger complex networks including examples taken from the real WWW.
From our numerical simulations we found  that  quantum PageRank has a series of unexpected properties.
In particular, we observed that the quantum algorithm, when applied on scale-free networks, is able to highlight the structure of the secondary hubs and to resolve the degeneracy in importance of the low-lying part of the list of rankings, which represents a typical shortcoming of the classical PageRank algorithm. When applied to hierarchical graphs,
the algorithm has the capability to better reveal the hierarchy of levels, of which the graph is composed, and to highlight the difference in importance of nodes within every hierarchy layer better than its classical PageRank counterpart, thus 
better resolving the connectivity structure of the underlying graph. 
Furthermore, the robustness of the quantum algorithm was analysed and the algorithm was shown to be less prone to significant changes in its output when the damping parameter was varied. The classical algorithm's output was known to  depend very strongly on this parameter. We found instead that the dependency in the quantum case was less significant than in the classical PageRank protocol. This finding indicates that the precise value of this parameter, whose choice is to some extent arbitrary, turns out to be not crucial for the quantum algorithm to work reliably.

Moreover, we have found that the distribution of importance values of quantum PageRanks over  the nodes of scale-free networks follow a power-law behaviour. A similar behaviour is found also for the classical PageRank algorithm.  However, the corresponding scaling exponent is for the quantum protocol smaller than in the classical case, indicating a smoother ranking of nodes and a more homogenous distribution of importance. In contrast to the classical algorithm, in the quantum protocol the hubs of the graphs do not concentrate the whole importance and the algorithm lifts the degeneracy of the large set of nodes with low importance values. This increased ranking capability comes at the cost of being more sensitive to structural changes to the network such as coordinate attacks on hubs.

Remarkably, the described characteristics of the quantum PageRank even persist if the algorithm is applied to real-world networks. We have studied and successfully tested the performance of the algorithm by applying it to a real-world network, originating from the hyperlink structure of www.epa.org \cite{Batagelj_Mrvar_2006}. Our study confirmed  that the intriguing properties of the quantum algorithm are not restricted to artificially, numerically grown networks.

Regarding the quantum PageRank algorithm as a directed quantum walk, we have studied the localisation properties of the quantum walker in the quantum protocol. An analysis of the Inverse Participation Ratio (IPR) in Ref.~\cite{paparo2013quantum} 
revealed that the quantum walker is localised in the case of the quantum PageRank applied to scale-free networks under standard conditions  (damping parameter $\alpha = 0.85 $).
This finding is consistent with the ability of the quantum algorithm to highlight hubs of the network.

The classical PageRank algorithm has been the subject of exact studies yielding analytical results~\cite{marijuan11}, and other studies have concentrated on properties of quantum walks on networks~\cite{Boettcher_2013,caruso2013universally,makmal2013quantum,Venegas-Andraca,faccin2013degree,kollar2014discrete,Whitfield} or adiabatic quantum computations of the classical PageRank~\cite{Lidar11}. 
Similarly, it is desirable to obtain exact analytical results on the quantum algorithm, which would complement the understanding gained from numerical studies, such as the one of Ref.~\cite{paparo2013quantum}.

In future work it will be interesting to analyse in more detail the impact of random failures of nodes in large networks of differing topology. Furthermore, from an algorithmic point of view, it is an interesting task to develop a dissipative version of this algorithm and to understand its performance and robustness properties in such scenario. Dissipation has already been considered as an element with respect to some aspect of the algorithm  \cite{Garnerone2012,zueco2012}, but the development of a truly dissipative version in the spirit of dissipative quantum algorithms and computation \cite{Verstraete2009,Diehl2009} remains an open question. Furthermore, the growing field of complex quantum networks would benefit from a version of the algorithm that is able to rank nodes in the more general case where qubits are located at the nodes of the network. An important question in this scenario is whether an algorithm based on a multi-particle quantum walk \cite{Childs2013,Childs2009} is needed in this context, or if there exists for this task an efficiently simulatable algorithm that belongs to the computational complexity class $P$.

\noindent {\em Acknowledgements.} This work has been supported by the
Spanish MINECO grants, the European Regional Development Fund
under projects  FIS2012-33152, MTM2011-28800-C02-01, CAM research
consortium QUITEMAD S2009-ESP-1594, European
Commission PICC: FP7 2007-2013, Grant No. 249958 and UCM-BS grant GICC-910758 and the U.S. Army Research Office through grant W911NF-14-1-0103.

\bibliographystyle{naturemag}
\bibliography{Bibliography}
\end{document}